\documentclass[11pt,sort&compress]{elsarticle}
\usepackage[paper=letterpaper,top=35mm, bottom=35mm, left=35mm, right=35mm]{geometry}
\makeatletter
\def\ps@pprintTitle{%
 \let\@oddhead\@empty
 \let\@evenhead\@empty
 \def\@oddfoot{\centerline{\thepage}}%
 \let\@evenfoot\@oddfoot}
\makeatother
\usepackage{amsmath}
\usepackage{accents}
\usepackage{amssymb}
\usepackage{mathrsfs}
\usepackage[LGR,T1]{fontenc}
\usepackage[latin1]{inputenc}
\let\oldbibliography\thebibliography
\renewcommand{\thebibliography}[1]{%
  \oldbibliography{#1}%
  \setlength{\itemsep}{1.4pt}%
}
\usepackage[cal=boondox,calscaled=1]{mathalfa}
\DeclareMathAlphabet{\bbvar}{U}{BOONDOX-ds}{m}{n}
\DeclareMathAlphabet{\bbgreek}{U}{bbold}{m}{n}
\usepackage[defaultsans,osfigures,scale=0.9]{opensans}
\usepackage{graphicx}
\usepackage{psfrag}
\usepackage{pstool}
\usepackage{caption}
\usepackage{graphicx}%
\newcommand{\hook}{\text{\large{$\lrcorner$}}}

\usepackage[colorlinks=true,
            linkcolor=blue,
            urlcolor=blue,
            citecolor=blue]{hyperref}
\usepackage[hyperref]{xcolor}
\definecolor{darkred}{rgb}{.95,.0,.0}
\hypersetup{
  colorlinks=true,
}

\newcommand{\di}{\mathrm{d}}
\usepackage{tensor}\newcommand{\ou}[3]{{#1}{}^{#2}{}_{#3}}
\newcommand{\uo}[3]{{#1}{}_{#2}{}^{#3}}
\newcommand{\vt}[2]{\tensor[^{\mathnormal{#1}}]{#2}{}}
\newcommand{\vo}[3]{\tensor[^{\mathnormal{#1}}]{#2}{^{#3}}}
\newcommand{\vu}[3]{\tensor[^{\mathnormal{#1}}]{#2}{_{#3}}}
\newcommand{\vou}[4]{\tensor[^{\mathnormal{#1}}]{#2}{^{#3}_{#4}}}
\newcommand{\vuo}[4]{\tensor[^{\mathnormal{#1}}]{#2}{_{#3}^{#4}}}

\newcommand{\I}{\mathrm{i}} 
\newcommand{\E}{\mathrm{e}} 
\newcommand{\CC}{\mathrm{cc.}} 
\newcommand{\C}{\mathbb{C}}

\newcommand{\R}{\mathbb{R}}

\newenvironment{subalign}{\subequations\align}{\endalign\endsubequations}
\newcommand{\eref}[1]{(\ref{#1})}

\DeclareMathAlphabet{\sfit}{OT1}{fos}{sb}{it}
\DeclareMathAlphabet{\mathsf}{OT1}{fos}{sb}{n}

\definecolor{darkgreen}{rgb}{0.01, 0.75, 0.24}

\usepackage[multiple, flushmargin]{footmisc}

\begin{document}

\begin{abstract} 
A field theory on a three-dimensional manifold is introduced, whose field equations are the constraint equations for general relativity on a three-dimensional null hypersurface. The underlying boundary action consists of two copies of the dressed Chern\,--\,Simons term for self-dual Ashtekar variables, a kinetic term for the null flag at the boundary plus additional junction conditions for the spin coefficients across the interface. In fact, there is a doubling of the field content, because the null hypersurface will be considered as an internal boundary between two adjacent slabs of spacetime. 
 The paper concludes with a proposal for a construction of the gravitational transition amplitudes in the bulk via the auxiliary boundary field theory alone, namely by gluing amplitudes for edge states across two-dimensional corners, thus providing a proposal for a quasi-local realisation of the holographic principle at the light front.
\end{abstract}%

\title{Generating functional for gravitational null initial data}
\author{Wolfgang Wieland}
\address{Perimeter Institute for Theoretical Physics\\31 Caroline Street North\\ Waterloo, ON N2L\,2Y5, Canada\\{\vspace{0.5em}\normalfont May 2019}
}
\maketitle
\vspace{-1.2em}
\hypersetup{
  linkcolor=black,
  urlcolor=black,
  citecolor=black
}
{\tableofcontents}
\hypersetup{
  linkcolor=darkred,
  urlcolor=red,
  citecolor=darkred
}
\begin{center}{\noindent\rule{\linewidth}{0.4pt}}\end{center}\newpage
\section{Introduction}
\noindent Recent developments have put to the forefront the role of gravitational boundary symmetries \cite{Balachandran:1994up,Wald:1999wa} in the context of the black hole information paradox, \cite{Hawking:2016msc,Hawking:2016sgy}. The corresponding boundary charges are defined on the event horizon of a black hole. The definition of a black hole horizon is highly non-local, and it can be argued, therefore, that these boundary charges \cite{Balachandran:1994up,Hawking:2016msc,Hawking:2016sgy,Wald:1999wa,Chandrasekaran:2018aop,Afshar:2016wfy} arise quite generally on all null surface boundaries: at null infinity as the generators of BMS transformations \cite{Bondi21,Sachs103,Barnich:2011mi}, but they also arise at generic null surfaces, as recently stressed in the context of non-perturbative quantum gravity \cite{Donnelly:2016auv,Wieland:2017zkf,Freidel:2016bxd}. 

The question is then whether the dynamics of these boundary charges can be understood holographically \cite{Bousso:2002ju}, without any reference to the bulk by strictly restricting ourselves to an auxiliary field theory intrinsic to the null boundary. The purpose of this paper is to demonstrate that such a boundary field theory exists for vacuum general relativity in four dimensions and non-vanishing cosmological constant $\Lambda$. The construction is based on earlier ideas in the context of non-perturbative quantum gravity and spinfoams \cite{wieland:nulldefects, Freidel:2018pbr,Freidel:2018pvm}, where a curved manifold is constructed by successively gluing patches of spacetime across boundaries. In fact, our basic strategy is to construct the boundary field theory at a null surface by gluing two adjacent slabs of spacetime, in each one of which the full non-linear $\Lambda$-vacuum Einstein equations are satisfied.\footnote{No restriction on the algebraic structure of the Weyl tensor at the boundary will be imposed. An approach to quasi-local holography for algebraically special solutions of Petrov type $D$ can be found in e.g.\ \cite{Goswami:2015zju}.} 
 At the null boundary, the fundamental configuration variables are therefore doubled. The underlying action has a very simple fundamental structure. In fact, it describes a prototypical example of a timeless system \cite{rovelli,PhysRevD.42.2638}, which consists of two composite systems on either side of the interface plus additional matching conditions for the quasi-local charges across the boundary. Schematically, the boundary action assumes the following form
\begin{equation}
S[p^\pm,q^\pm,N^I]=\int\Big(p_i^+\di q^i_+-p_i^-\di q^i_--\di t\, N^I\big(H_I(p_+,q_+)-H_I(p_-,q_-)\big)\Big),
\end{equation}
where $N^I$ are Lagrange multipliers imposing gluing conditions for the boundary charges. What were constraint equations for general relativity on a null hypersurface before, will now turn into evolution equations for gravitational corner data (edge modes) along the null generators. The propagating fields of the three-dimensional boundary theory are the gravitational edge modes alone, the two radiative modes (bulk gravitons) appear as external sources in the action.

The paper is divided into three parts: the first part, \hyperref[sec2]{section 2}, introduces the kinematical framework, which is based on the constraint equations for general relativity in terms of self-dual Ashtekar variables \cite{newvariables,ashtekar} on a null hypersurface. The second part, \hyperref[sec3]{section 3}, develops the boundary field theory and demonstrates that the resulting boundary field equations are nothing but the constraint equations for general relativity on a null boundary. The third part, \hyperref[sec4]{section 4} concludes the paper and provides an outlook for what are the next steps ahead: the basic idea is to define the quantum states for general relativity in the bulk via the auxiliary boundary field theory. Transition amplitudes between edge states at two consecutive cross-sections of the null boundary will depend parametrically on external sources, namely the gravitational flux that may cross the null surface in between. A boundary transition amplitude defines, therefore, a complex-valued functional on an extended boundary state space, which now captures both the radiative modes along the null surface (as external sources) as well as gravitational edge states at the two-dimensional corners. The distinction between states and amplitudes is therefore a matter of perspective: what represents an amplitude from the perspective of the boundary field theory defines a state on a three-dimensional portion of a null hypersurface. This viewpoint clearly resonates with the general boundary formalism \cite{Oeckl:2005bv,rovelli} and more recent developments in loop quantum gravity, where a quasi-local realisation of the holographic principle has been developed within non-perturbative and three-dimensional quantum gravity, see \cite{Wieland:2018ymr,Dittrich:2017hnl,Dittrich:2018xuk,Bonzom:2015ans,Asante:2018kfo}.

A few remarks regarding our conventions. In the following, we will be working on an oriented three-manifold $\mathcal{N}$ of topology $\mathcal{N}\simeq[0,1]\times S^2$. If reference to a four-dimensional spacetime manifold $\mathcal{M}$ is needed, $\mathcal{N}$ will be part of its boundary: $\mathcal{N}\subset\partial\mathcal{M}$. Most of our considerations are, however, strictly intrinsic to $\mathcal{N}$, and reference to the bulk is only made to identify the field content and its boundary dynamics. In addition, we will use the abstract index convention: three and four-dimensional tensors will carry abstract co(tangent) indices $a,b,c,\dots$. If there is a chance of confusion, we will use the prefix $4$ to distinguish differential forms in the bulk from fields intrinsic to the null surface. Greek indices $\alpha,\beta,\dots$ will denote internal Lorentz indices with respect to a (co)tetrad $\vou{4}{e}{\alpha}{a}$ in $\mathcal{M}$. Upper case indices $A, A', B,B',\dots$ will denote $SL(2,\C)$ spinors in the spin $(\tfrac{1}{2},0)$ resp.\ complex conjugate spin $(0,\tfrac{1}{2})$ representation. The metric signature is $(-$$+$$+$$+)$. Lorentz vectors $V^{\alpha}$ are therefore indentified with \emph{anti-hermitian} $(\tfrac{1}{2},\frac{1}{2})$ spinors $V^{AA'}=-\bar{V}^{A'A}$. Spinor indices are raised and lowered with the skew-symmetric $\epsilon_{AB}$-tensor and its inverse, i.e.\ $\epsilon^{AB}:\epsilon^{AC}\epsilon_{BC}=\delta^A_B\equiv\uo{\epsilon}{A}{B}$, such that $\xi^A=\epsilon^{AB}\xi_B$ and $\xi_A\eta^A=-\eta^A\xi_A$. A final comment on the status of the soldering forms and internal vectors from the perspective of the space of histories: the skew-symmetric $\epsilon_{AB}$-tensors as well as the internal Minkowski metric $\eta_{\alpha\beta}$ will be treated as internal background structures, whose field variations vanish, while the dynamical fields of the theory consist of the soldering forms $\ou{e}{\alpha}{a}$ and the spin connection coefficients $\ou{A}{\alpha}{\beta a}$ at the null hypersurface. 
\section{Ashtekar variables on a null surface}\label{sec2}
\subsection{Spinors, dyads, triads, tetrads on the null boundary\label{sec2.1}}\noindent
On a spacelike hypersurface $\Sigma$, there is a canonical projector $\ou{h}{a}{b}=\ou{g}{a}{b}+n^an_b$ and a canonical normal direction $n^a:n^an_a=-1$ with respect to which we can always decompose tensor fields $\ou{T}{a\dots}{b\dots}\in T^{r}_{s}\mathcal{M}$ into their respective space and time components. On a null hypersurface $\mathcal{N}$, no such decomposition is available, because the normal direction lies now within the null hypersurface itself. A useful way to still speak about the boundary intrinsic geometry is to restrict attention to Lorentz-valued differential forms $\vo{4}{\omega}{\alpha\beta\dots}{}_{[ab\dots]}$, for which there still exists a natural notion of projection, namely the pull-back, which does not require a metric. Consider, for example, the pull-back of the cotetrad $\vou{4}{e}{\alpha}{a}$ to the null boundary $\mathcal{N}$,
\begin{equation}
\varphi^\ast_{\mathcal{N}}\vou{4}{e}{\alpha}{a}=\ou{e}{\alpha}{a}.\label{nulltetra1}
\end{equation}
Any such cotetrad on $\mathcal{N}$ can be then parametrised in terms of a cotangent space triad $(k_a,m_a,\bar{m}_a)\in T^\ast\mathcal{N}$ and a pair of spinors that parametrise the internal directions,
\begin{equation}
\ou{e}{\alpha}{a}\equiv \ou{e}{AA'}{a}=-\I\ell^A\bar{\ell}^{A'}k_a+\I \ell^A \bar{k}^{A'}m_a+\I k^A\bar{\ell}^{A'}\bar{m}_a,\label{nulltetra2}
\end{equation}
where we introduced an $\epsilon$-normalised basis $(k^A, \ell^A):\epsilon_{AB}k^A\ell^B=1$ in the spin $(\frac{1}{2},0)$ frame bundle over $\mathcal{N}$, and implicitly identified \emph{internal} Lorentz vector fields with sections of the anti-hermitian spin $(\tfrac{1}{2},\tfrac{1}{2})$ spin bundle over the boundary, see e.g.\ \cite{penroserindler} for the details of the notation. 

The dyadic component $(m_a,\bar{m}_a)$ plays an important role in the following. It provides sufficient data to reconstruct \emph{both} the induced signature $(0\,+$$+)$ three-metric,
\begin{equation}
\varphi^\ast_{\mathcal{N}}\vu{4}{g}{ab} =:q_{ab}=2m_{(a}\bar{m}_{b)},\label{2-metrc}
\end{equation}
as well as the canonical area two-form on $\mathcal{N}$,
\begin{equation}
\varepsilon_{ab}=2\I m_{[a}\bar{m}_{b]}.\label{area2-form}
\end{equation}
Notice also that the boundary dyad $(m_a,\bar{m}_a)$ is charged under an internal $U(1)$ gauge symmetry: given the pull-back $q_{ab}$ of the spacetime metric $\vu{4}{g}{ab}$ alone, the one-forms $(m_a,\bar{m}_a)$ are unique up to the $U(1)$ gauge transformations
\begin{equation}
m_a \longrightarrow \E^{\I\phi}m_a\qquad\E^{\I \phi}:\mathcal{N}\rightarrow U(1).\label{U1gauge}
\end{equation}
For any given such triadic cobasis $(k_a,m_a,\bar{m}_a)$ on $\mathcal{N}$, we then also have a dual tangent space basis $(\ell^a,m^a,\bar{m}^a)$, whose elements satisfy
\begin{equation}
\ell^ak_a=-1,\quad \ell^am_a=0,\quad m^am_a=0,\quad\bar{m}^am_a=1.
\end{equation}
The vector $\ell^a\in T\mathcal{N}$ is a null generator of $\mathcal{N}$ and lies tangent to the null generators of the null hypersurface $\mathcal{N}$. Since there is no preferred normalisation for $\ell^a$, we must also consider the dilatations,
\begin{equation}
\ell^a \longrightarrow \E^\omega\ell^a,\qquad\omega:\mathcal{N}\rightarrow\R.\label{dilat}
\end{equation}
On field space, the direction of these null generators will be considered as a universal structure shared between different spacetimes. The ruling of $\mathcal{N}$ is therefore seen as a fiducial background structure, which can be identified with the kernel of the complex-valued one-form $m_a\in T^\ast_\C\mathcal{N}$.

The dyadic one-form $m_a$ is canonical on $\mathcal{N}$: given the geometry in the bulk, it is unique only up to the residual $U(1)$ gauge transformations \eref{U1gauge}. The one-form $k_a$, on the other hand, depends on the foliation. If the null boundary $\mathcal{N}$ carries a preferred time orientation, as we assume in here, we can always choose a time function $u:\mathcal{N}\rightarrow\R$ such that $l^a\partial_a u> 0$ for any future pointing null generator $l^a\sim \ell^a$. We may then foliate $\mathcal{N}$ into $u=\mathrm{const}.$ hypersurfaces $S^2_u\simeq\{u\}\times S^2$.  Given such a fiducial foliation, we then also have a decomposition of the boundary one-form $k_a$ into the  lapse and shift functions $\Phi:\mathcal{N}\rightarrow \R_>$ and $N:\mathcal{N}\rightarrow\C$,
\begin{equation}
k_a=-\Phi\partial_au+\bar{N}m_a+N\bar{m}_a.\label{lapseshift}
\end{equation}
From the perspective of the boundary field theory, we will treat the spin dyad $(k^A,\ell^A)$ as a dynamical field at the boundary. It is then always possible to introduce a trivial field redefinition to absorb the lapse and shift functions into the spin dyad: the transformation 
\begin{subalign}
\ell^A&\longrightarrow \ell^{A}_\ast=\sqrt{\Phi}\ell^A,\label{shiftl}\\
k^A&\longrightarrow k^A_\ast=\frac{1}{\sqrt{\Phi}}\left(k^A-N\ell^A\right),\label{shiftk}
\end{subalign}
preserves our canonical normalisation, i.e.\ $\epsilon_{AB}k^A\ell^B=\epsilon_{AB}k^A_\ast\ell^B_\ast=1$ and defines an \emph{internal} $SL(2,\C)$ Lorentz transformation that maps $(k^A,\ell^A)$ into $(k^A_\ast,\ell^A_\ast)$. Such a field redefinition brings the boundary tetrad \eref{nulltetra2} into the following simplified form,
\begin{equation}
\ou{e}{\alpha}{a}\equiv \ou{e}{AA'}{a}=\I\ell^A\bar{\ell}^{A'}\partial_au+\I \ell^A \bar{k}^{A'} m_a+\I k^A\bar{\ell}^{A'}\bar{m}_a.\label{nulltetra3}
\end{equation}
Hence
\begin{equation}
k_a=-\partial_au\label{kdef}
\end{equation}
without loss of generality. Notice that this procedure is different from fixing lapse and shift by introducing supplementary gauge conditions, by demanding, for example, that $\ell^a$ be affinely parametrised, because we have merely reabsorbed the lapse and shift functions $\Phi$ and $N$ into the spin dyad $(k^A,\ell^A)$ without imposing any conditions on the connection coefficients.

To conclude this brief introduction, consider the following spinor-valued two-form,
\begin{equation}
\eta_A = \big(\ell_A k-k_A m)\wedge \bar{m},\label{etadef}
\end{equation}
which satisfies
\begin{equation}
\eta_{Aab}\ell_B=\frac{1}{2\I}\varepsilon_{ab}\epsilon_{AB}+\Sigma_{ABab},
\end{equation}
where $\Sigma_{ABab}$ denotes the self-dual  Pleba\'nski two-form on the null hypersurface, which is defined as the self-dual part of the bivector $e_\alpha\wedge e_\beta$,
\begin{equation}
\Sigma_{AB}=-\frac{1}{2}e_{AC'}\wedge \uo{e}{B}{C'}=\varphi^\ast_{\mathcal{N}}\vu{4}{\Sigma}{AB}.\label{Sigmadef}
\end{equation}

The two-form $\eta_{Aab}$ plays an important role for the quasi-local Hamiltonian analysis on a null surface. Given the  symplectic structure in the bulk, the pull back of $\eta_{Aab}$ to a two-dimensional $u=u_o$ cross-section $S^2_o$ of the null boundary is the canonical conjugate momentum to the \emph{null flag} $\ell^A$, such that the boundary spinors $\ell^A$ and $\pi_A:=1/(8\pi\I G)\times\varphi^\ast_{S^2}\eta_{Aab}$ generate an infinite-dimensional Heisenberg algebra on the cross-section of the null hypersurface \cite{Wieland:2017zkf,Wieland:2017cmf}. 

\subsection{Ashtekar connection on a null boundary}
\noindent On a null hypersurface, and in the absence of additional universal structures, such as a preferred foliation, symmetries, restricted boundary conditions such as those imposed by isolated horizons or null infinity, there is no preferred metric-compatible, torsionless covariant derivative intrinsic to a null hypersurface \cite{AshtekarNullInfinity,Ashtekar:aa}. In the general case, the only available boundary connection is the one inherited by taking the pull-back from the bulk.  At the level of the spin bundle, the resulting boundary spin connection is the generalisation of the self-dual Ashtekar connection\footnote{On a spacelike hypersurface  $\Sigma$, the real part $\ou{\Gamma}{i}{a}$ of the Ashtekar connection denotes the intrinsic $\mathfrak{su}(2)$ spin connection with respect to a cotriad $\ou{e}{i}{a}$ on $\Sigma$, while $K_{ab}=\ou{K}{i}{b}e_{ia}$ is the extrinsic curvature.}
\begin{equation}
\ou{A}{i}{a}=\ou{\Gamma}{i}{a}+\I\ou{K}{i}{a}\label{Ashdef}
\end{equation}
from spacelike hypersurfaces to null hypersurfaces. The only difference to the spacelike case is that there is now no unique way to split $\ou{A}{i}{a}$ into real and imaginary parts, simply because there is no unique reference connection $\ou{\Gamma}{i}{a}$ with respect to which we could decompose $\ou{A}{i}{a}$ into extrinsic and intrinsic components.

Given a connection, we also have a covariant derivative: if $\vo{4}{\psi}{A}$ denotes a smooth spinor-field in the vicinity of the null boundary $\mathcal{N}$, and if ${\psi}^A$ denotes the restriction of this field to the boundary, the covariant derivative $D_a$ is simply given by
\begin{equation}
D_a\psi^A=\varphi^\ast_{\mathcal{N}}\nabla_a\vo{4}{\psi}{A},\quad \text{for}\quad\nabla_a:2\nabla_{[a}\vo{4}{e}{\alpha}{}_{b]}=0.
\end{equation}
In fact, if we want to describe initial data for general relativity on a null hypersurface, the Ashtekar connection must be compatible with the pull-back of the torsionless condition,\footnote{In this paper, we are only considering vacuum general relativity. In the presence of half-integer spin fields, and depending on the coupling to gravity, the spin density of fermions may appear on the right hand side of this equation.}
\begin{equation}
2D_{[a} \ou{e}{AA'}{b]}=0.\label{torsless}
\end{equation}

To understand how this condition imposes restrictions on the affine space of connections on the null hypersurface, it is useful to consider first the anholonomy coefficients of the triadic basis: for any generic null cotriad $(k_a,m_a,\bar{m}_a)$ on $\mathcal{N}$, we introduce the exterior derivatives and define the corresponding component functions,
\begin{subalign}
\di m&=-\frac{1}{2}\Big(\vartheta_{(\ell)}+2\I\varphi\Big)k\wedge m-\sigma_{(\ell)}k\wedge\bar{m}+\I\gamma m\wedge\bar{m},\label{m-exd}\\
\di k&=\lambda k\wedge\bar{m}+\bar{\lambda} k\wedge m+\I\tau_{(k)} m\wedge \bar{m},\label{k-exd}
\end{subalign}
where $\di$ denotes the ordinary exterior derivative on $\mathcal{N}$. All coefficients have a clear geometric interpretation: $\sigma_{(\ell)}$ and $\vartheta_{(\ell)}$ are the shear and expansion of the null generators $T\mathcal{N}\ni\ell^a:k_a\ell^a=-1$, whereas $\omega_a=\varphi k_a+\bar{\gamma}m_a+\gamma \bar{m}_a$ defines a boundary intrinsic $U(1)$ connection; $\tau_{(k)}$, on the other hand, is the twist of the transversal (and internal) null direction\footnote{If we only know the pull-back of the cotetrad to the null surface, namely \eref{nulltetra1}, then $k^\alpha=\I k^A\bar{k}^{A'}$ only exists as an \emph{internal} null vector, because we do not have the inverse tetrad $\vuo{4}{e}{\alpha}{a}$ at hand to map $k^\alpha$ into $k^a=k^\alpha\vuo{4}{e}{\alpha}{a}\in T\mathcal{M}$.} $k^{AA'}=\I k^A\bar{k}^{A'}$ and $\lambda$ is the $\ell^a$-component of the Lie bracket $[\ell,m]^a=\ell^b\partial_b m^a-m^b\partial_b \ell^a=\mathcal{L}_\ell m^a$.

The torsionless conditions impose $4 \times 3=12$ constraints on the Ashtekar connection on $\mathcal{N}$, which has $6\times 3=18$ independent algebraic components. By imposing the torsionless conditions $De^{AA'}=0$, there remain, therefore,  six unconstrained spin coefficients, which  are a measure for the extrinsic curvature of the null hypersurface. To see how these six unconstrained coefficients lie within the Ashtekar connection on $\mathcal{N}$, consider first the difference tensor,
\begin{equation}
\Delta^{AB}:=k^{(A}D\ell^{B)}-\ell^{(A}Dk^{B)},\label{Deltadef1} 
\end{equation}
that satisfies
\begin{equation}
D_ak^A=\ou{\Delta}{A}{Ba}k^B,\qquad D_a\ell^A=\ou{\Delta}{A}{Ba}\ell^B.\label{Deltadef2} 
\end{equation}
Given the exterior derivatives (\ref{m-exd}, \ref{k-exd}) of the triadic cobasis $(k_a,m_a,\bar{m}_a)$ on $\mathcal{N}$, it is straight-forward to show that the torsionless condition \eref{torsless} on a null hypersurface restricts a generic such difference tensor $\ou{\Delta}{A}{Ba}$ to the following algebraic form,
\begin{align}
\nonumber\ou{\Delta}{AB}{a}=&-\Big[(\kappa+\I\varphi)k_a+\I\big(\bar{\gamma}+\I(\bar\lambda-\bar\alpha)\big)m_a+\I\big({\gamma}+\I(\lambda-\alpha)\big)\bar{m}_a\Big]\ell^{(A}k^{B)}+\\
\nonumber&-\Big[\frac{1}{2}\big(\vartheta_{(k)}-\I\tau_{(k)}\big)\bar{m}_a+\bar{\sigma}_{(k)}m_a+\bar{\alpha} k_a\Big]\ell^A\ell^B+\\
&+\Big[\frac{1}{2}\vartheta_{(\ell)}m_a+\sigma_{(\ell)}\bar{m}_a\Big]k^Ak^B.\label{difftensor}
\end{align}
The only unspecified components of the Ashtekar connection on a null boundary are therefore the transversal shear $\sigma_{(k)}$, the transerval expansion $\vartheta_{(k)}$, the acceleration (non-affinity) $\kappa$ and the additional spin coefficient $\alpha$. All other components are fixed by the boundary intrinsic exterior derivatives \eref{m-exd} and \eref{k-exd}. The acceleration $\kappa$ and the transversal expansion $\vartheta_{(k)}$ are real-valued functions on $\mathcal{N}$, the shear $\sigma_{(k)}$ an the spin coefficient $\alpha$ are complex-valued. This leaves us with six unspecified degrees of freedom $(\kappa,\vartheta_{(k)},\sigma_{(k)},\alpha)$ per point on $\mathcal{N}$ that we interpret, in analogy to the spacelike case, as a measure for the extrinsic curvature, which is now no longer a boundary intrinsic tensor, as in the spacelike case, but merely a collection of certain preferred spin coefficients on the null hypersurface.\footnote{Under dilatations \eref{dilat}, the non-affinity $\kappa$ transforms like the $u$-component of an abelian connection.}

If we want to impose the torsionless condition \eref{torsless} at the level of a boundary field theory, it will prove useful to decompose the constraint into its irreducible spin components. For instance, the traceless and symmetric spin $(1,1)$ tensor contribution is given by
\begin{equation}
\ou{X}{AB}{A'B'}=\ou{e}{(A}{A'}\wedge D\ou{e}{B)}{B'}=0.\label{11torsion}
\end{equation}
The spin $(1,0)$ bivector constraint, on the other hand, is
\begin{equation}
X_{AB}=e_{AC'}\wedge D \uo{e}{B}{C'}=D\Sigma_{AB},\label{10torsion}
\end{equation}
which is nothing but the exterior covariant derivative of the self-dual Pleba\'nski two-form \eref{Sigmadef}. There is also the scalar component $e_{AA'}\wedge De^{AA'}$, but it is redundant. In fact, the system of constraints (\ref{11torsion}, \ref{10torsion}) is equivalent to \eref{torsless}. This can be shown explicitly by comparing the individual components, but it is way more instructive to compare the number of algebraically independent constraints directly: the spin $(1,0)$ constraint $X_{AB}=0$ imposes three complex conditions, the constraint $X_{ABA'B'}$, on the other hand, satisfies $X_{ABA'B'}=X_{(AB)(A'B')}=\bar{X}_{A'B'AB}$ and\footnote{This happens because of $\I\ell_A\bar{\ell}_{A'}e^{AA'}=\varphi^\ast_{\mathcal{N}}\ell_\alpha\vou{4}{e}{\alpha}{a}=\varphi^\ast_{\mathcal{N}}\ell_a=0$.} $X_{ABA'B'}\ell^A\bar{\ell}^{A'}\ell^B\bar{\ell}^{B'}=0$, which give $9-1=8$ independent real conditions. In addition, the components  $X_{AB}\ell^A\ell^B$ and $X_{ABA'B'}\ell^A\ell^B\bar\ell^{A'}\bar{k}^{B'}$ are linearly dependent,
\begin{subalign}
X_{AB}\ell^A\ell^B&=-\I\bar{m}\wedge\ell^A\bar{\ell}^{A'}D e_{AA'}\stackrel{!}{=}0,\\
X_{ABA'B'}\ell^A\ell^B\bar\ell^{A'}\bar{k}^{B'}&=-\I\bar{m}\wedge\ell^A\bar{\ell}^{A'}D e_{AA'}\stackrel{!}{=}0,
\end{subalign}
such that $X_{AB}=0$ and $X_{ABA'B'}=0$ together impose $6+8-2=12$ real and linearly independent constraints, which are obtained linearly from $De^\alpha=0$ on $\mathcal{N}$, which are $4\times 3=12$ constraints as well. In summary, imposing \eref{11torsion} in addition to \eref{10torsion} is the same as to say that the pull-back to the null surface of the torsion two-form vanishes.

Notice also that $X_{ABA'B'}$ has the same algebraic structure as the traceless part of the Ricci tensor in the bulk. This observation is an important hint for how to construct the gravitational boundary field theory on a null hypersurface $\mathcal{N}$. We will see, in fact, that Lagrange multipliers that impose constraints on the connection coefficients \eref{difftensor} will turn into sources for the pull-back of the curvature tensor to a null hypersurface. 

\subsection{Field strength, constraints, evolution equations}
\noindent Before introducing the boundary field theory, let us first recall the algebraic structure of the Riemann curvature tensor for a generic solution of the $\Lambda$-vacuum Einstein equations on a null hypersurface \cite{penroserindler}. Consider first the decomposition of the four-dimensional Riemann curvature tensor into its irreducible components, 
\begin{subalign}
\vou{4}{R}{\alpha}{\beta cd}&=\delta^{A'}_{B'}\vou{4}{F}{A}{Bcd}+\delta^{A}_{B}\vou{4}{\bar{F}}{A'}{B'cd},\\
\vou{4}{F}{A}{Bcd}&=\frac{R}{12}\vou{4}{\Sigma}{A}{Bcd}+\ou{\Psi}{AB}{CD}\vou{4}{\Sigma}{CD}{ab}+\ou{\Phi}{AB}{A'B'}\bar\Sigma^{A'B'},
\end{subalign}
where $R=\bar{R}$ is the Ricci scalar, $\Phi_{ABA'B'}=\Phi_{(AB)(A'B')}=\bar{\Phi}_{ABA'B'}$ is the traceless part of the Ricci tensor and $\Psi_{ABCD}=\Psi_{(ABCD)}$ denotes the spin $(2,0)$ Weyl spinor. The $\Lambda$-vacuum Einstein equations imply
\begin{equation}
R=4\Lambda,\qquad\Phi_{ABA'B'}=0.
\end{equation}
Taking the pull-back to $\mathcal{N}$, we obtain the curvature of the selfdual Ashtekar connection on the null boundary,
\begin{equation}
F^{AB}=\frac{\Lambda}{3}\eta^{(A}\ell^{B)}+\ou{\psi}{AB}{C}\eta^C,\label{Fbndry}
\end{equation}
where $\psi_{ABC}=\psi_{(ABC)}=\Psi_{ABCD}\ell^D$. Next, we define the boundary intrinsic components of the Weyl tensor, namely
\begin{subalign}
\Psi_4&=\psi_{ABC}\ell^A\ell^B\ell^C,\label{psi4}\\
\Psi_3&=\psi_{ABC}\ell^A\ell^Bk^C,\label{psi3}\\
\Psi_2&=\psi_{ABC}\ell^Ak^Bk^C,\label{psi2}\\
\Psi_1&=\psi_{ABC}k^Ak^Bk^C.\label{psi1}
\end{subalign}
Using the definition of the curvature tensor on $\mathcal{N}$, namely
\begin{equation}
F^{AB}=D\Delta^{AB}-\ou{\Delta}{A}{C}\wedge \Delta^{CB},
\end{equation}
we may then express the spin curvature components (\ref{psi4}--\ref{psi1}) in terms of the connection coefficients \eref{difftensor}. There are two sets of equations. First of all, we have the constraint equations,
\begin{subalign}
\nonumber \Psi_1 = &\frac{1}{2}\mathcal{L}_{\bar m}\big(\vartheta_{(k)}-\I\tau_{(k)}\big)-\frac{1}{2}\bar\beta(\vartheta_{(k)}-\I\tau_{(k)})+\\\label{psi1eq}
&\hspace{9em}-\mathcal{L}_m\bar{\sigma}_{(k)}+\I(2\gamma-\I\beta)\bar{\sigma}_{(k)}+\I\bar\alpha\tau_{(k)},\\\label{psi2eq}
\nonumber\Psi_2=&\frac{\Lambda}{6}-\frac{1}{2}\Big[\I\mathcal{L}_{\bar{m}}\big(\gamma-\I\beta\big)-\bar\gamma\big(\gamma-\I\beta\big)-\I\mathcal{L}_m\big(\bar\gamma-\I\bar\beta\big)-\gamma(\bar\gamma-\I\bar\beta)+\\
&\hspace{9em}+\I\tau_{(k)}(\kappa-\tfrac{1}{2}\vartheta_{(\ell)}+\I\varphi)+\tfrac{1}{2}\vartheta_{(k)}\vartheta_{(\ell)}-2\sigma_{(\ell)}\bar{\sigma}_{(k)}\Big],\\\label{psi3eq}
\Psi_3=&\frac{1}{2}\big(\mathcal{L}_m\vartheta_{(\ell)}+\beta\vartheta_{(\ell)}\big)-\mathcal{L}_{\bar m}\sigma_{(\ell)}-\I(2\bar\gamma-\I\bar\beta)\sigma_{(\ell)}.
\end{subalign}
where $\mathcal{L}_V$ is the Lie derivative for e.g. $V^a=m^a\in T\mathcal{N}$ and $\beta$ denotes the abbreviation
\begin{equation}
\beta:=\alpha-\lambda,
\end{equation}
for spin coefficients $\alpha$ and $\lambda$ defined as in \eref{difftensor} and \eref{k-exd}. On the other hand, there are also the boundary intrinsic evolution equations
\begin{subalign}
\mathcal{L}_\ell&\sigma_{(\ell)}+\big(\vartheta_{(\ell)}-\kappa-2\I\varphi\big)\sigma_{(\ell)}=-\Psi_4,\label{evolv1}\\
\mathcal{L}_\ell&\vartheta_{(\ell)}+\big(\tfrac{1}{2}\vartheta_{(\ell)}-\kappa\big)\vartheta_{(\ell)}+2\sigma_{(\ell)}\bar{\sigma}_{(\ell)}=0\label{evolv2}\\
\mathcal{L}_{\ell}&\bar{\sigma}_{(k)}+(\kappa+\tfrac{1}{2}\vartheta_{(\ell)}+2\I\varphi)\bar{\sigma}_{(k)}=-\mathcal{L}_{\bar{m}}\bar\alpha+\I\bar\gamma\bar\alpha+\bar\alpha\bar\alpha-\tfrac{1}{2}(\vartheta_{(k)}-\I\tau_{(k)})\bar{\sigma}_{(\ell)},\label{evolv3}\\\nonumber
\mathcal{L}_\ell&\big(\vartheta_{(k)}-\I\tau_{(k)}\big)+\big(\kappa+\tfrac{1}{2}\vartheta_{(\ell)}\big)\big(\vartheta_{(k)}-\I\tau_{(k)}\big)=\\\label{evolv4}
&\hspace{6em}=-2\mathcal{L}_m\bar\alpha+2\I\gamma\bar\alpha+2\alpha\bar\alpha-2\sigma_{(\ell)}\bar\sigma_{(k)}+\frac{2\Lambda}{3}+2\Psi_2,\\
\mathcal{L}_\ell&\big(\bar\gamma-\I\bar\beta\big)+\big(\tfrac{1}{2}\vartheta_{(\ell)}+\I\varphi\big)\big(\bar\gamma-\I\bar\beta\big)+\bar{\sigma}_{(\ell)}(\gamma-\I\beta)=\nonumber\\\label{evolv5}
&\hspace{6em}=\I\mathcal{L}_{\bar{m}}\big(\kappa+\I\varphi\big)+\I\bar\lambda\big(\kappa+\I\varphi\big)+\I\bar\alpha\vartheta_{(\ell)},\\
\nonumber\mathcal{L}_\ell&\big(\gamma-\I\beta\big)+\big(\tfrac{1}{2}\vartheta_{(\ell)}-\I\varphi\big)\big(\gamma-\I\beta\big)+\sigma_{(\ell)}(\bar\gamma-\I\bar\beta)=\\
\label{evolv6}&\hspace{6em}=\I\mathcal{L}_m\big(\kappa+\I\varphi\big)+\I\lambda(\kappa+\I\varphi)+2\I\bar\alpha\sigma_{(\ell)}+2\I\Psi_3.
\end{subalign}
The set of equations (\ref{psi1eq}--\ref{psi3eq}) and (\ref{evolv1}--\ref{evolv6}) is underdetermined: the gravitational flux that may cross the null hypersurface is unconstrained and may be encoded into $\Psi_4$, which must be characterised as an external datum. Besides the flux, additional gauge fixing conditions for $\kappa$, $\lambda$, $\tau_{(k)}$ and $\varphi$ must be chosen to integrate the evolution equations along the null generators $\ell^a$. A possible choice is e.g.\ $\varphi=0$, $\lambda=0$, $\tau_{(k)}=0$ and $\kappa=\tfrac{1}{2}\vartheta_{(\ell)}$. Notice also that \eref{evolv5}, \eref{evolv6} and \eref{psi3eq} are linearly dependent, which follows from the integrability condition derived from $\di^2\,m=0$.

The remaining Einstein equations that contain the off-boundary components of the curvature two-form and torsion two-form, namely $\vou{4}{F}{A}{Bab}k^b$ and $\vou{4}{T}{AA'}{ab}k^b$ for a transversal null direction $k^a\in T\mathcal{M}:\vu{4}{g}{ab}\ell^a k^b=-1$ have an altogether different status from the perspective of the boundary field theory: they turn into transversal evolution equations\footnote{In Bondi coordinates, the transversal evolution equations turn into the radial evolution equations with respect to the affine Bondi $r$-coordinate.} that continue the boundary intrinsic fields into the bulk. In here, we are only interested in the field theory intrinsic to the boundary, and do not consider the transversal evolution equations, because we regard these hypersurface deformations as gauge. This viewpoint is motivated by background invariance, which implies that the location of the boundary with respect to a given coordinate system has no physical meaning, and is justified by the fact that a transversal vector field $\xi^a\in T\mathcal{M}:\xi^a|_{\mathcal{N}}\notin T\mathcal{N}$ of compact support on the null boundary generates a field-variation $\delta_\xi[\cdot]=\mathcal{L}_\xi(\cdot)$ on the covariant phase space that lies in the kernel of the pre-symplectic two-form \cite{Wald:1999wa}.

\section{Generating functional and boundary field theory}\label{sec3}
\subsection{What is to be fixed at the boundary?}\label{sec3.1}
\noindent The purpose of this paper is to present a generating functional for null initial data\,---\,an action for a field theory on the three-manifold $\mathcal{N}$, whose equations of motion are the constraint equations for gravitational initial data on a partial\footnote{For definiteness, we assume that the null boundary has a definite topology $\mathcal{N}=[0,1]\times S^2$ with boundaries $S^2_-=\{0\}\times S^2$ and $S^2_{+}=\{1\}\times S^2$. } and light-like Cauchy hypersurface $\mathcal{N}$. Since we are working on a partial Cauchy surface, such a three-dimensional boundary field theory must necessarily depend on additional sources that characterise the missing data, e.g.\ gravitational radiation that might cross the null boundary.\footnote{To integrate the initial data on an initial $u=\mathrm{const}.$ cross-section along the null generators, additional gauge fixing conditions for the spin coefficients $\kappa$, $\lambda$, $\tau_{(k)}$ and $\varphi$ must be chosen.}  The most immediate way to specify this data is to treat the radiative component of the Weyl tensor on $\mathcal{N}$, namely 
\begin{equation}
\Psi_4=\Psi_{ABCD}\ell^A\ell^B\ell^C\ell^D
\end{equation}
as an external source at the boundary. Instead of fixing $\Psi_4$, we may, however, also just fix its potential, namely the shear $\sigma_{(\ell)}$ of the null generators $\ell^a$, see \eref{evolv1}. Since the shear $\sigma_{(\ell)}$ is obtained, in turn, from the Lie derivative $\mathcal{L}_\ell m =\ell\hook(\di m)$ of the dyad, we may remove yet another $\ell^a$-derivative and impose yet another class of boundary conditions, namely require conditions on the allowed field variations of the dyadic basis $(m_a,\bar{m}_a)$. This is, in fact, Sachs's original choice \cite{Sachs:1962zzb}. 

Following this logic, we introduce restrictions on the allowed field variations at the boundary. Consider first the definition of lapse and shift on the null hypersurface, as given in \eref{lapseshift}. We have seen in equation \eref{shiftl} and \eref{shiftk} that the lapse and shift functions can be reabsorbed always into the definition of the spin dyad $(k^A,\ell^A)$. But now that the dyad is taken itself as an independent field at the boundary, whose field variations are subject only to the normalisation condition,
\begin{equation}
k_A\delta\ell^A-\ell_A\delta k^A=0,
\end{equation}
we may always absorb variations of lapse and shift via \eref{shiftl} and \eref{shiftk} into variations of the spin dyad. We can therefore always assume that the variations of $k_a$ vanish, in other words
\begin{equation}
\delta k_a=0,
\end{equation}
without loss of generality. This in turn allows us to align $k_a$ with a fiducial foliation on $\mathcal{N}$, and we thus say,
\begin{equation}
k_a=-\partial_a u\in T^\ast \mathcal{N}.\label{kfixed}
\end{equation}
Since we now also have $\di k=0$, the spin coefficients $\lambda=\alpha-\beta$ and $\tau_{(k)}$ vanish, which simplifies the Ashtekar connection \eref{difftensor} and the constraint  and evolution equations, (\ref{psi1eq}--\ref{psi3eq}) and (\ref{evolv1}--\ref{evolv6}) on the null hypersurface.

We are now left to specify the variations of the dyadic one-forms $(m_a,\bar{m}_a)$. A general variation of $m_a$ on $\mathcal{N}$ can be decomposed again into the triadic basis $(k_a,m_a,\bar{m}_a)$,
\begin{equation}
\delta m_a=f m_a + \varsigma \bar{m}_a+\zeta k_a, 
\end{equation}
Sachs's choice \cite{Sachs:1962zzb} is essentially a conformal gauge,\footnote{Analogous gauge conditions also exist in euclidean gravity, see e.g.\ \cite{Wieland:2018ymr,Witten:2018lgb} for recent applications.} where the off-diagonal variations $\varsigma$ and $\zeta$ are set to zero, and we are left with only conformal variations
\begin{equation}
\delta m_a=f m_a.\label{mvar}
\end{equation}
What is kept fixed as an external source at the boundary is therefore an equivalence class of complex-valued one-forms,
\begin{equation}
h:=[m_a]\ni m_a^\prime\Leftrightarrow \exists\,\E^{f}:\mathcal{N}\rightarrow \C: m_a^\prime=\E^{f}m_a,\label{mclass}
\end{equation}
namely the conformal two-structure at the boundary \cite{Sachs:1962zzb,InvernoStachelConformal}. The conformal boundary conditions can be therefore summarised as
\begin{equation}
\delta h =0.\label{boundrycond}
\end{equation}

Let us now briefly recall the prescription to integrate the evolution equations along the  null generators $\ell^a$. First of all, we have to fix supplementary gauge conditions on the diagonal $\ell^a$-components of the self-dual connection \eref{difftensor} at the boundary. A possible choice is
\begin{subalign}
\kappa&=\frac{1}{2}\vartheta_{(\ell)},\label{gaugecond1}\\
\varphi &= 0.\label{gaugecond2}
\end{subalign}
Next, and in order to integrate the equations of motion (\ref{evolv1}--\ref{evolv6}), we pick a fiducial representative $\vu{\circ}{m}{a}$ in the equivalence class of dyadic one-forms $m_a$,
\begin{equation}
m^\circ_a\in[m_a]\Leftrightarrow \exists \Omega:m_a = \Omega\,m^\circ_a,
\end{equation}
where $\Omega:\mathcal{N}\rightarrow\C$ is the conformal factor. Consider then the exterior derivative of the fiducial one-form $\vu{\circ}{m}{a}\in T^\ast\mathcal{N}$. We introduce a decomposition as in \eref{m-exd} and write
\begin{align}
\di m^\circ &= -\frac{1}{2}\Big({\vartheta}^\circ_{(\ell)}+2\I\varphi^\circ\Big)k\wedge m^\circ-\sigma_{(\ell)}^\circ k\wedge m^\circ+\I\gamma^\circ\,m^\circ\wedge\bar{m}^\circ.
\end{align}
Comparison with \eref{m-exd} provides,\footnote{Since $(\ell^a,\vo{\circ}{m}{a},\vo{\circ}{\bar{m}}{a})=(\ell^a,\Omega^{-1}m^a,\Omega^{-1}\bar{m}^a)$ is the $T\mathcal{N}$ tangent space basis dual to $(k_a,m_a^\circ,\bar{m}_a^\circ)$, the Lie derivatives $\mathcal{L}_m$ can expressed in terms of the fiducial derivatives  $\mathcal{L}_{\vt{\circ}{m}}$ and $\mathcal{L}_{\ell}$ alone.}
\begin{subalign}
\gamma&=\I\Omega^{-1}\mathcal{L}_m\Omega+\bar{\Omega}^{-1}\gamma^\circ,\label{Omegatrans1}\\
\sigma_{(\ell)}&={\bar\Omega}^{-1}{\Omega}\,\sigma^\circ_{(\ell)},\label{Omegatrans2}\\
\tfrac{1}{2}\vartheta_{(\ell)}&=\Omega^{-1}\mathcal{L}_\ell\Omega+\tfrac{1}{2}\vartheta^\circ_{(\ell)}+\I(\varphi^\circ-\I\varphi).\label{Omegatrans3}
\end{subalign}

The prescription to integrate the boundary equations of motion (\ref{evolv1}--\ref{evolv6}) along the null generators can be now summarised as follows, see \cite{Sachs:1962zzb} and \cite{Torre:1985rw,DePaoli:2017sar,Reisenberger:2018xkn} for a $2+2$ Hamiltonian approach to the problem. Consider first \eref{Omegatrans2}, which implies that the  shear $\sigma_{(\ell)}$ is completely fixed by the  shear $\sigma^\circ_{(\ell)}$ of the fiducial one-form $m_a^\circ$ and the phase of $\Omega$. The $U(1)$ connection $\gamma$, on the other hand, is  determined by the spatial derivatives of the conformal factor, see \eref{Omegatrans1}. If we go then back to the constraint equation \eref{psi2eq} for $\Psi_2$ and also take into account the gauge conditions \eref{gaugecond1} in addition to $\lambda=0$ and $\tau_{(k)}=0$, which are a consequence of $\di k=0$, we immediately see that the system of evolution equations \eref{Omegatrans3}, \eref{evolv2}, \eref{evolv3} and \eref{evolv5} defines a system of first-order differential equations for $\Omega$, $\vartheta_{(\ell)}$, $\vartheta_{(k)}$, $\sigma_{(k)}$ and $\alpha$, which determine the so far unconstrained spin coefficients, see \eref{difftensor}. 
The initial data (the gravitational edge modes, so to say) on an $u=u_o$ cross-section of the null boundary are therefore given by the value of the conformal factor $\Omega|_{u_o}$, the boundary intrinsic  expansion $\vartheta_{(\ell)}|_{u_o}$, the transversal shear and expansion, $\sigma_{(k)}|_{u_o}$ and $\vartheta_{(k)}|_{u_o}$, plus the value of the additional spin coefficient $\alpha|_{u_o}$ at a two-dimensional $u=u_o$ initial cross-section of the null boundary.

So far, we have merely discussed and reorganised past results. The next step ahead is to present an action $S[p^\pm,q^\pm|h]$ for certain dynamical fields $p^\pm,q^\pm$ and fixed external sources $h=[m_a]:\delta h=0$ such that the resulting boundary field equations return us back the constraint equations (\ref{evolv1}--\ref{evolv6}) and (\ref{psi1eq}--\ref{psi3eq}) of general relativity on a null hypersurface. In this way, the constraint equations of general relativity on a null surface turn into actual dynamical field equations, thereby realising a quasi-local version of the holographic principle at the light front.
 
\subsection{Kinematical structure}
\noindent The key idea of the proposal is to double the field content. We will treat, in fact, the null boundary $\mathcal{N}$ as an interface between two adjacent slabs $\mathcal{M}^+$ and $\mathcal{M}^-$ of spacetime, in each one of which the $\Lambda$-vacuum Einstein equations are satisfied.\footnote{The assumption is that the $\Lambda$-vacuum Einstein equations  $R_{ab}-\tfrac{1}{2}Rg_{ab}+\Lambda g_{ab}=0$ are satisfied in some neighbourhood of the common null interface.} The question is then how to glue the two slabs together without introducing new and unphysical degrees of freedom at the interface. On top of that, we will only consider those $\Lambda$-vacuum geometries, where the  $\Psi_4$, $\Psi_3$, $\Psi_2$ and $\Psi_1$ components of the Weyl tensor are all continuous across the interface, thus excluding, unfortunately, many interesting solutions of impulsive gravitational waves \cite{Aichelburg1971,Penrose:1972xrn,Griffiths:1991zp}, such as e.g.\ the Penrose\,--\,Khan geometry \cite{Khan:1971aa}.

Having doubled the field content, we now have two independent $SL(2,\C)$ principal bundles $P_{\pm}(SL(2,\C),\mathcal{N})$ and associate spin bundles on either side of the interface. Our fundamental configuration variables of the boundary field theory are therefore given by the $SL(2,\C)$ Ash\-te\-kar connection on either side, the dyadic frame fields and the triadic basis: a point $\mathcal{h}$ in the space of histories is given by field configurations
\begin{equation}
\mathcal{h}=\big(\ou{[A_+]}{A}{B},(k_+^A,\ell^A_+),(k^+_a,m^+_a);\ou{[A_-]}{A}{B},(k_-^A,\ell^A_-),(k^-_a,m^-_a)\big)\label{histspace}
\end{equation}
on $\mathcal{N}$. The key idea to construct the boundary field theory is then to impose matching conditions across the interface. Our first condition concerns the metric degrees of freedom. The assumption is that the entire intrinsic geometry is matched across the null hypersurface,\footnote{These conditions can be weakened. Depending on whether we only match the area two-form \eref{area2-form} or the entire three-metric \eref{2-metrc}, we would generate  twisted geometries \cite{freidelsimotwist} or ordinary impulsive gravitational waves \cite{Aichelburg1971,Griffiths:1991zp,Penrose:1972xrn}. In this paper, we do not consider such distributional configurations.}
\begin{subalign}
m^+_a&=m^-_a=m_a,\label{gluecond1}\\
k^+_a&=k^-_a=k_a=-\partial_a u\label{gluecond2}.
\end{subalign}
Given the triadic basis $(k_a,m_a,\bar{m}_a)$ and the spin dyads $(k^A_\pm,\ell^A_\pm)$ on $\mathcal{N}$, we have a hierarchy of composite fields on the null hypersurface. There are the tetrads, the spinor-valued two-forms \eref{etadef}, and the self-dual Pleba\'nski two-form on $\mathcal{N}$, namely,
\begin{subalign}
e^{AA'}_\pm & = \I\ell^A_\pm\bar{\ell}^{A'}_\pm\di u+\I \ell^A_\pm \bar{k}^{A'}_\pm m+\I k^A_\pm\bar{\ell}^{A'}_\pm m,\label{pmtetra}\\
\eta_A^\pm & = -\big(\ell_A^\pm \di u+k_A^\pm m\big)\wedge \bar{m}\label{pmeta}\\
\Sigma^\pm_{AB} & = \eta^\pm_{(A}\ell^\pm_{B)},\label{pmSigma}
\end{subalign}
for normalised spin dyads\footnote{Geometrically speaking, there are two distinguished such epsilon tensors $\epsilon^\pm_{AB}$ in either frame, but we can drop the $\pm$ suffix without the possibility of confusion. For the same reason, we write $D^\pm \ell^A\equiv D\ell^A_\pm=\di \ell^A+\ou{[A_\pm]}{A}{B}\ell^B_\pm$.} $\epsilon_{AB}k^A_+\ell^B_+=\epsilon_{AB}k^A_-\ell^B_-=1$.

\subsection{Kinetic term}
\noindent Before introducing the boundary field theory in the next section, and as warm up for our construction, consider first the self-dual action for gravity in the bulk \cite{selfdualtwo,Plebanski:1977zz}, which is given by 
\begin{equation}
S[\vt{4}{\Sigma},\vt{4}{A},\Psi]=\frac{\I}{8\pi G}\int_{\mathcal{M}}\Big[\vou{4}{\Sigma}{AB}\wedge \vo{4}{F}{AB}-\frac{1}{2}\Psi_{ABCD}\vo{4}{\Sigma}{AB}\wedge\vo{4}{\Sigma}{CD}-\frac{\Lambda}{6}\vu{4}{\Sigma}{AB}\wedge\vo{4}{\Sigma}{AB}\Big],
\end{equation}
where the Weyl spinor $\Psi_{ABCD}$ serves as a Lagrange multiplier for the simplicity constraints $\vu{4}{\Sigma}{(AB}\wedge\vu{4}{\Sigma}{CD)}=0$. On a null hypersurface, and for the usual Dirichlet boundary conditions we need an analogue of the Gibbons\,--\,Hawking boundary term to cancel the connection variation from the bulk. In terms of complex variables, this boundary term is given by the integral \cite{Wieland:2017zkf}
\begin{equation}
S[\eta_A,\ell^A|A]=\frac{\I}{8\pi G}\int_{\mathcal{N}}\eta_A\wedge D\ell^A.\label{bndryterm}
\end{equation}
This term is essentially a symplectic structure for our boundary fields $(\eta^\pm_A,\ell^A_\pm)$. Indeed, the natural kinetic term for the boundary field theory is simply the integral,
\begin{equation}
S_{\text{kin}}[(k_\pm,\ell_\pm),A_\pm,m|h]=\frac{\I}{8\pi G}\int_{\mathcal{N}}\Big[\eta^+_A\wedge D\ell^A_+-\eta^-_A\wedge D\ell^A_-\Big].\label{kinterm}
\end{equation}
The notation $S[\dots,m|h]$ indicates that we only take such variations of $m_a$ into account that satisfy \eref{mvar} and treat the equivalence classes $h=[m_a]$ as an external and fixed source, see \eref{boundrycond}.

\subsection{Dressed Chern\,--\,Simons term}
\noindent At the null boundary, we will treat both the connection as well as the boundary spinors $(k^A,\ell^A)$ as dynamical variables. An action, which realises this principle, is given by the dressed $SL(2,\C)$ Chern\,--\,Simons functional,
\begin{equation}
S_{\mathrm{CS}}[k^A,\ell^A,\ou{A}{A}{Ba}]=\frac{1}{2}\int_{\mathcal{N}}\mathrm{Tr}\big[F\wedge\Delta-\frac{1}{3}\Delta\wedge\Delta\wedge\Delta\big],\label{CSdressed}
\end{equation}
where the $\mathfrak{sl}(2,\C)$ trace is defined as $\mathrm{Tr}(FX)=\ou{F}{A}{B}\ou{X}{A}{B}=-F^{AB}X_{AB}$ and $\ou{\Delta}{A}{Ba}$ is the difference tensor \eref{difftensor} for the  normalised spin dyad $(k^A,\ell^A):\epsilon_{AB}k^A\ell^B=1$. The variation of this action with respect to the spin dyad vanishes up to a boundary term thanks to the Bianchi identity $DF_{AB}=0$. Variations of the $SL(2,\C)$ connection, on the other hand, return us back the curvature plus a boundary term,
\begin{equation}
\delta_{A}S_{\mathrm{CS}}[k^A,\ell^A,\ou{A}{A}{Ba}]=+\frac{1}{2}\int_{\partial\mathcal{N}}\Delta_{AB}\wedge\delta A^{AB}-\int_{\mathcal{N}}F_{AB}\wedge\delta A^{AB}.
\end{equation}

\subsection{Boundary action and boundary field theory}
\noindent In this section, we will develop the boundary action, whose equations of motion are the evolution and constraint equations of four-dimensional vacuum general relativity on a null hypersurface, namely (\ref{psi1eq}--\ref{psi3eq}) and (\ref{evolv1}--\ref{evolv5}). The basic idea is to couple the kinetic term \eref{kinterm} for the boundary spinors to the dressed Chern\,--\,Simons functional \eref{CSdressed} and impose additional gluing and gauge-fixing conditions. In fact, already by adding the two terms \eref{kinterm} and \eref{CSdressed} alone, we are left with an action, which is  a generating functional for the self-dual part of the De Sitter curvature $\mathcal{F}_{\alpha\beta}=F_{\alpha\beta}-\frac{\Lambda}{3}e_\alpha\wedge e_\beta$ on a null hypersurface,
\begin{equation}
\delta_{A_\pm} \Big[S_{\mathrm{kin}}+\frac{3\I}{8\pi\Lambda G}S_{\mathrm{CS}}\Big]=\pm\frac{3\I}{8\pi\Lambda G}\int_{\mathcal{N}}\Big(F^\pm_{AB}-\frac{\Lambda}{3}\eta^\pm_{(A}\ell^{\pm}_{B)}\Big)\wedge\delta A^{AB}_{\pm}.\label{test1}
\end{equation}
As it stands, this action is, however, defective: the resulting field equations are way too restrictive, because the Weyl spinors $\Psi_1$, $\Psi_2$, $\Psi_3$ and $\Psi_4$ would all vanish on $\mathcal{N}$. There would be no gravitational radiation crossing the null hypersurface nor static solutions in the bulk. What makes matters even worse is that the action \eref{test1} violates the torsionless conditions \eref{torsless}. In fact, only one half of them would be imposed by \eref{test1}, namely the spin $(1,0)$ components $D\Sigma^{\pm}_{AB}=0$, which follow from the Bianchi identities $DF^{\pm}_{AB}=0$ and $\mathcal{F}_{\alpha\beta}=0$. 

To improve \eref{test1} and turn it into a generating functional for $\Lambda$-vacuum general relativity on a null hypersurface, additional gluing conditions have to be imposed: so far, we only matched the intrinsic metric geometries across the null interface, namely \eref{gluecond1} and \eref{gluecond2}. The connection coefficients are still completely uncorrelated. 
Accordingly, two sets of constraint equations will be imposed, namely for both the boundary intrinsic connection coefficients $(\varphi,\gamma,\lambda,\tau_{(k)},\vartheta_{(\ell)},\sigma_{(\ell)})$ and the extrinsic curvature components $(\kappa,\vartheta_{(k)},\sigma_{(k)},\alpha)$, see \eref{difftensor}. First of all, we introduce additional spin $(1,0)$ and spin $(1,1)$ Lagrange multipliers $B^\pm_{AB}$ and $B^\pm_{ABA'B'}=B^\pm_{(AB)(A'B')}=\bar{B}^\pm_{A'B'AB}$ to impose the torsionless conditions \eref{11torsion} and \eref{10torsion}. Since we have already matched the intrinsic metric geometry across the null interface, the imposition of the torsionless condition will then also imply matching constraints for the boundary intrinsic spin coefficients,
\begin{equation}
\varphi^+=\varphi^-,\quad\gamma^+=\gamma^-,\quad\vartheta^+_{(\ell)}=\vartheta^-_{(\ell)},\quad\sigma^+_{(\ell)}=\sigma^-_{(\ell)},\quad
\tau_{(k)}^\pm=0,\quad\lambda^\pm=0,\label{connctnmatch1}
\end{equation}
where e.g.\ $\gamma^\pm$ is the spatial $\bar{m}_a$ component of the diagonal $U(1)$ part of the difference tensor $\vou{\pm}{\Delta}{AB}{a}$, which is defined in above \eref{difftensor}. Notice also that $\tau_{(k)}$ and $\lambda^{\pm}$ vanish thanks to our gauge conditions, namely $k^+_a=k^-_a=-\partial_au$, see \eref{gluecond2}.

Next, we have to impose that also the extrinsic components of the spin connection are matched across the interface. To impose these constraints in a gauge invariant fashion, consider first the transition function that brings us from one spin frame into the other, 
\begin{equation}
\ou{U}{A}{B}=\ell_+^Ak^-_A-k_+^A\ell^-_A\in SL(2,\C).\label{Udef}
\end{equation}
Going back to the defining properties of the difference tensor \eref{Deltadef2}, we immediately find
\begin{equation}
D\ou{U}{A}{B}=\di \ou{U}{A}{B}+\vou{+}{A}{A}{C}\ou{U}{C}{B}-\ou{U}{A}{C}\vou{-}{A}{C}{B}=\vou{+}{\Delta}{A}{C}\ou{U}{C}{B}-\ou{U}{A}{C}\vou{-}{\Delta}{C}{B}.\label{DU}
\end{equation}
To match the extrinsic components $\kappa^\pm$, $\sigma^\pm_{(k)}$ and $\alpha^\pm$ across the null hypersurface $\mathcal{N}$, consider the following spin $(\tfrac{3}{2},0)$ constraint,\footnote{The matching of $\vartheta_{(k)}$ will then follow from the matching of the initial conditions $\vartheta^+_{(k)}|_{u_o}=\vartheta^-_{(k)}|_{u_o}$ and the evolution equation \eref{evolv4}.}
\begin{equation}
C_{ABC}=\eta^+_{(A}\wedge[UD U^{-1}]_{BC)}=-\eta^+_{(A}\wedge\Delta^+_{BC)}+\eta^-_{(A}\wedge\Delta^-_{BC)}=0.\label{matchcond}
\end{equation}
This constraint has four independent components, namely
\begin{subalign}
C_3&=\ell^A_+\ell^B_+\ell^C_+C_{ABC}=m\wedge\bar{m}\wedge\big(\ell^+_AD\ell^A_+-\ell^-_AD\ell^A_-\big),\label{matchcond1}\\
C_2&=k^A_+\ell^B_+\ell^C_+C_{ABC}=\frac{1}{3}k\wedge\bar{m}\wedge\ell^+_AD\ell^A_++\frac{2}{3} m\wedge\bar{m}\wedge k^+_AD\ell^A_+-(+\leftrightarrow -),\label{matchcond2}\\
C_1&=k^A_+k^B_+\ell^C_+C_{ABC}=\frac{1}{3}m\wedge \bar{m}\wedge k^+_ADk^A_++\frac{2}{3}k\wedge\bar{m}\wedge k^+_AD\ell^A_+-(+\leftrightarrow -),\label{matchcond3}\\
C_0&=k^A_+k^B_+k^C_+C_{ABC}=k\wedge\bar{m}\wedge k^+_ADk^A_+-k\wedge\bar{m}\wedge k^-_ADk^A_-.\label{matchcond4}
\end{subalign}
If the torsionless conditions \eref{10torsion} and \eref{11torsion} are satisfied, we can use the decomposition of the difference tensor \eref{difftensor} to immediately see that $C_3$ vanishes by itself,\footnote{The null hypersurface is ruled by lightlike geodesics, the null generators $\ell^a$ are tangent to them, hence $\ell_A\ell^aD_a\ell^A=0$.} $\ell_A\ell^aD_a\ell^A=0$, while the three remaining constraints impose the matching conditions
\begin{subalign}
C_2=0&\Leftrightarrow \kappa^+=\kappa^-,\label{match1}\\
C_1=0&\Leftrightarrow {\bar{\alpha}}^+={\bar{\alpha}}^-,\label{match2}\\
C_0=0&\Leftrightarrow {\bar{\sigma}}^+_{(k)}={\bar{\sigma}}^-_{(k)}.\label{match3}
\end{subalign}

At this point, we are now only left to specify the gauge conditions to integrate the spin coefficients along the null generators. Our choice will be the following,
\begin{subalign}
\kappa^+=\kappa^-=\frac{1}{2}\vartheta_{(\ell)},\label{gaugecond1}\\ 
\varphi^+=\varphi^-=0.\label{gaugecond2}
\end{subalign}
To complete the construction of the action, we introduce additional Lagrange multipliers and add the matching conditions \eref{matchcond}, the torsionless conditions (\ref{10torsion}, \ref{11torsion}) and the previously mentioned gauge conditions to the generating functional \eref{test1} for the self-dual part of the De Sitter curvature. The resulting gauge-fixed action is therefore given by
\begin{align}
\nonumber S_{\mathrm{gf}}\big[(k^A_\pm,\ell^A_\pm),\vou{\pm}{A}{A}{Ba}&,\psi_{ABC},m_a,B^{\pm}_{\alpha\beta},B,\tilde{\xi}\big|h\big]=\\
\nonumber=\frac{3\I}{8\pi\Lambda G}\int_{\mathcal{N}}\bigg[&+\frac{\Lambda}{3}\eta^+_A\wedge D\ell^A_+ -\frac{1}{2}\mathrm{Tr}\big[F^+\wedge\Delta^+-\frac{1}{3}\Delta^+\wedge\Delta^+\wedge\Delta^+\big]+
 \\
&\nonumber- \frac{\Lambda}{3}\eta^-_A\wedge D\ell^A_-+\frac{1}{2}\mathrm{Tr}\big[F^-\wedge\Delta^--\frac{1}{3}\Delta^-\wedge\Delta^-\wedge\Delta^-\big]+\\
&\nonumber+\psi_{ABC}\,\eta^A_+\wedge\big[UDU^{-1}\big]^{BC}
+\\
&\nonumber +B^{+}_{\alpha\beta}e^\alpha_+\wedge D e^\beta_+-B^{-}_{\alpha\beta}e^\alpha_-\wedge D e^\beta_-+\\
&-3B\big(\ell^+_{(A}\ell^+_Bk^+_{C)}\eta^A_+\wedge\Delta^{BC}_+-\ell^-_{(A}\ell^-_Bk^-_{C)}\eta^A_-\wedge\Delta^{BC}_-\big)+\frac{3}{2}\tilde{\xi}B^2\bigg].\label{gfixedactn}
\end{align}
This action consists of three parts that all play a different role. The first two lines are the generating functional for the De Sitter curvature on either side of the null hypersurface. The third term is the matching condition for the extrinsic curvature components, and finally, there are the terms that involve the auxiliary $B$-fields: the $B_{\alpha\beta}$-terms impose the torsionless conditions \eref{torsless}, whereas the gauge fixing conditions for the spin coefficients $\kappa$ and $\varphi$, see \eref{difftensor}, are imposed via the auxiliary $B$-field. The variation of the scalar density $\tilde{\xi}$ yields the supplementary gauge condition $B=0$.

It will prove useful to split the $B^\pm_{\alpha\beta}$-variations into irreducible components. The variation of the traceless part is redundant and can be dropped: as explained in above,  the spin $(1,0)$ and spin $(1,1)$ components \eref{10torsion} and \eref{11torsion} are already enough to impose the pull-back of the torsionless equation at the null boundary. Accordingly, we may always assume that $B_{\alpha\beta}$ is traceless, and introduce the decomposition,
\begin{equation}
B_{\alpha\beta}=-\epsilon_{AB} \bar{B}_{A'B'}-\bar\epsilon_{A'B'}B_{AB}+C_{ABA'B'}.
\end{equation}
The variations of $B_{AB}=B_{BA}$ impose the spin $(1,0)$ component \eref{10torsion} of the torsionless equation on the null hypersurface, while the Lagrange multiplier $C_{ABA'B'}=C_{(AB)(A'B')}=\bar{C}_{A'B'AB}$ belongs to the spin $(1,1)$ contribution \eref{11torsion}, since
\begin{equation}
B_{\alpha\beta}e^\alpha\wedge De^\beta = -\bar{B}_{A'B'}D\bar{\Sigma}^{A'B'}-B_{AB}D\Sigma^{AB}+C_{ABA'B'}e^{AA'}\wedge D e^{BB'},\label{BeDecomponents}
\end{equation}
which we will need in below.

\subsection{Boundary equations of motion}
\noindent Having defined the gauge-fixed action \eref{gfixedactn}, we must now consider the corresponding field equations and check whether they generate null initial data for $\Lambda$-vacuum general relativity on a partial and light-like Cauchy hypersurface.

A number of variations have already implicitly been performed: as seen from \eref{10torsion} and \eref{11torsion}, the variation of $B^\pm_{\alpha\beta}=-\epsilon_{AB} \bar{B}^\pm_{A'B'}-\bar\epsilon_{A'B'}B^\pm_{AB}+C^\pm_{ABA'B'}$ impose the pull-back of the torsionless equation on either side of the null interface, namely $De^\alpha_\pm=0$. The variation of the spin $(\tfrac{3}{2},0)$ Lagrange multiplier $\psi_{ABC}$, on the other hand, imposes the matching conditions (\ref{match1}--\ref{match3}) for the extrinsic spin coefficients, $\kappa^\pm$, $\bar\sigma_{(k)}^\pm$ and $\alpha^\pm$. There are therefore, only three remaining field variations to be considered: the variation of the connection coefficients, the variation of the normalised spin basis $(k_\pm^A,\ell_\pm^A)$ and the variation of the codyad $(m_a,\bar{m}_a)$ for given boundary conditions (\ref{gluecond1}, \ref{gluecond2}) and \eref{boundrycond}.

Consider first the variation of the Lorentz connection  $\vou{\pm}{A}{\alpha}{\beta}$, which splits into self-dual and anti-self-dual components,
\begin{equation}
\ou{A}{\alpha}{\beta a}=\delta^A_B\ou{A}{A}{Ba}+\delta^{A'}_{B'}\ou{A}{A}{Ba}.
\end{equation}
The action \eref{gfixedactn} is complex, and the anti-self-dual connection only appears linearly in the $B^\pm_{\alpha\beta}$-terms that impose the torsionless condition,
\begin{equation}
De^{AA'}=\di e^{AA'}+\ou{A}{A}{B}\wedge e^{BA'}+\ou{\bar A}{A'}{B'}\wedge e^{AB'}.
\end{equation}
It follows, therefore, that the variation of the action with respect to the anti-self-dual connection $\vou{\pm}{\bar{A}}{A'}{B'a}$ must be a linear function of the auxiliary $B^\pm_{\alpha\beta}$-fields. Indeed,
\begin{align}
\delta_{\bar{A}_\pm}S_{\mathrm{gf}}=\pm\frac{3\I}{8\pi\Lambda G}\int_{\mathcal{N}}\Big[C^\pm_{ABA'B}\Sigma_\pm^{AB}\wedge
+2\bar{B}^\pm_{C'(A'}\vou{\pm}{\bar\Sigma}{C'}{B')}\Big]\wedge\delta\bar{A}^{A'B'}_\pm.
\end{align}
The corresponding field equations impose an algebraic constraint on the components of the auxiliary $B^\pm_{\alpha\beta}$-field, namely
\begin{equation}
C^\pm_{ABA'B'}\Sigma^{AB}_\pm+2\bar{B}^{\pm}_{C'(A'}\vou{\pm}{\bar\Sigma}{C'}{B')}=0.
\end{equation}
Since $B^\pm_{\alpha\beta}$ is real-valued, this also implies
\begin{equation}
C^\pm_{ABA'B'}\bar{\Sigma}^{A'B'}_\pm+2{B}^{\pm}_{C(A}\vou{\pm}{\Sigma}{C}{B)}=0.\label{Bterms}
\end{equation}
Consider next the variation of the self-dual part of the connection. The $\vou{+}{A}{A}{B}$ variation can be immediately inferred from the variation of the generating functional, namely \eref{test1}, and the covariant derivative $DU=\di U +A_+ U- UA_-$ of the transition function appearing in the gluing conditions \eref{matchcond}. A short calculation yields
\begin{align}\nonumber
\delta_{A_+}S_{\mathrm{gf}}\approx\frac{3\I}{8\pi\Lambda G}\int_{\mathcal{N}}\bigg[&-\frac{\Lambda}{3}\eta^+_A\ell^+_B+F^+_{AB}+C^+_{CAC'D'}e_+^{CC'}\wedge\uo{[e_+]}{B}{D'}+\\
&+2B^+_{CA}\ou{[\Sigma^+]}{C}{B}-\Psi_{ABC}{\eta}_+^C\bigg]\wedge\delta A_+^{AB},\label{Aplusvar}
\end{align}
where $\approx$ denotes equality up to terms that vanish \emph{on-shell}. In the same way, we then also have
\begin{align}
\nonumber \delta_{A_-}S_{\mathrm{gf}}\approx\frac{3\I}{8\pi\Lambda G}\int_{\mathcal{N}}\bigg[&+\frac{\Lambda}{3}\eta^-_A\ell^-_B-F^-_{AB}-C^-_{CAC'B'}e_+^{CA'}\wedge\uo{[e_-]}{B}{C'}+\\
&-2B^-_{CA}\ou{[\Sigma^-]}{C}{B}+\Psi_{EFC}{\eta}_+^C\ou{U}{E}{A}\ou{U}{F}{B}\bigg]\wedge\delta A_-^{AB}.\label{Aminvar}
\end{align}
If we now go back to the definition of the self-dual Pleba\'nski two-form \eref{Sigmadef}, and insert back the equations of motion for the auxiliary $B$-field, namely \eref{Bterms}, we arrive at the boundary field equations
\begin{equation}
F^\pm_{AB}=\frac{\Lambda}{3}\eta^\pm_{(A}\ell^\pm_{B)}+\psi^\pm_{ABC}\eta^C_\pm,\label{bndryF}
\end{equation}
where we defined the boundary Weyl spinors on either side of the interface, namely
\begin{equation}
\psi^+_{ABC}:=\psi_{ABC},\quad\text{and}\quad \psi^-_{ABC}:=\psi_{EFD}\ou{U}{E}{A}\ou{U}{F}{B}\ou{U}{D}{C}.
\end{equation}

Since the connection is torsionless and the pair $(\eta_A^\pm,\ell^A_\pm)$ of boundary spinors encodes the entire intrinsic geometry of the null hypersurface, the resulting field equations \eref{bndryF} are equivalent to the constraint equations (\ref{psi1eq}--\ref{psi3eq}) and (\ref{evolv1}--\ref{evolv6}) of general relativity on a null hypersurface.

Before discussing these results in more detail, let us first complete the variation of the action with respect to the remaining boundary fields, namely the spin bases $(k^A_\pm,\ell^A_\pm)$ and the dyadic one-forms $(m_a,\bar{m}_a)$ for given conformal boundary conditions \eref{boundrycond}. The variation of the dyadic basis, $(k_+^A,\ell^A_+)$ can be performed as follows. Consider a gauge parameter $\ou{[\phi_+]}{A}{B}:\mathcal{N}\rightarrow\mathfrak{sl}(2,\C)_+$, and define a corresponding field variation $\delta_{\phi_+}$, whose only non-vanishing components are given by
\begin{subalign}
\delta_{\phi_+}[k^A_+]&=\ou{[\phi_+]}{A}{B}k^B_+,\\
\delta_{\phi_+}[\ell^A_+]&=\ou{[\phi_+]}{A}{B}\ell^B_+,\\
\delta_{\phi_+}[\psi_{ABC}]&=-3\psi_{F(AB}\ou{[\phi_+]}{F}{C)}.
\end{subalign}
All other field variations vanish, in particular $\delta_{\phi_+} A_\pm=0$. The dressed Chern\,--\,Simons functional is invariant under infinitesimal such frame rotations, and the only contribution to the $\delta_{\phi_+}$-variation of the gauge-fixed action is therefore given by,
\begin{equation}
\delta_{\phi_+} S_{\mathrm{gf}}\approx\frac{3\I}{8\pi\Lambda G}\int_{\mathcal{N}}\bigg[-\frac{\Lambda}{3}\eta^+_{(A}\ell^+_{B)}\wedge D\phi^{AB}_+-\psi_{ABC}\eta_+^C\wedge D\phi^{BC}_+\bigg],
\end{equation}
where $\approx$ denotes equality up to terms that are constrained to vanish provided the gauge fixing conditions \eref{gaugecond1} and \eref{gaugecond2} are satisfied. Since $\phi_{AB}^+$ vanishes at the boundary of $\mathcal{N}$ and the Bianchi identities hold on all of $\mathcal{N}$, i.e.\ $DF^+_{AB}=0$, the $\delta_{\phi_+}$-variation yields a total derivative and vanishes, therefore, on shell,
\begin{equation}
\delta_{\phi_+} S_{\mathrm{gf}}\approx 0.
\end{equation}
In the same way, one can then also demonstrate that the variations of the spin basis $(k^A_-,\ell^A_-)$ vanish: let $\ou{[\phi_-]}{A}{B}:\mathcal{N}\rightarrow\mathfrak{sl}(2,\C)_-$ gauge element that generates  an infinitesimal gauge transformation of the  $(k^A_-,\ell^A_-)$ frame fields,
\begin{subalign}
\delta_{\phi^-}[k^A_-]&=\ou{[\phi^-]}{A}{B}k^B_-,\\
\delta_{\phi^-}[\ell^A_-]&=\ou{[\phi^-]}{A}{B}\ell^B_-,
\end{subalign}
whereas all other field variations on the space of histories vanish. From the invariance of the Chern\,--\,Simons functional \eref{CSdressed} under small $SL(2,\C)$ gauge transformations  and thanks to $\phi^{AB}_-\big|_{\partial\mathcal{N}}=0$, we are then again left with
\begin{equation}
\delta_{\phi^-} S_{\mathrm{gf}}\approx\frac{3\I}{8\pi\Lambda G}\int_{\mathcal{N}}\bigg[+\frac{\Lambda}{3}\eta^-_{(A}\ell^-_{B)}\wedge D\phi^{AB}_-+\psi_{CEF}\eta_+^C\ou{U}{E}{A}\ou{U}{F}{B}\wedge D\phi^{AB}_+\bigg].
\end{equation}
As for the $\delta_{\phi^+}$-variations, the right hand side vanishes on shell: going back to the field equations \eref{bndryF}, we can replace the integrand by $F^-_{AB}\wedge D\phi^{AB}_-$, which is a total derivative, because of the Bianchi identities $DF^{AB}_-=0$.

We are now only left to consider the variations of the boundary dyad $m_a$ for given boundary conditions \eref{boundrycond}, which will be studied in the next section. 
\subsection{Conformal transformations}
\noindent Sachs's  boundary conditions \eref{boundrycond} restrict the field variations of the dyadic coframes $(m_a,\bar{m}_a)$ to conformal transformations,
\begin{equation}
\delta_fm_a=fm_a,\qquad f:\mathcal{N}\rightarrow\C,
\end{equation}
while $\delta_f k_a=-\delta_f\partial_a u=0$ without loss of generality see \hyperref[sec2.1]{section 2.1}. In the last section, we have already considered the field variations of all other configuration variables on the space of histories \eref{histspace}. We are therefore free to declare how the vector field $\delta_f$ on field space \eref{histspace} acts on the spin bases $(k^A_\pm,\ell^A_\pm)$. A particularly useful choice is given by
\begin{subalign}
\delta_f\ell^A_\pm&=+\frac{f}{2}\ell^A_\pm,\\
\delta_fk^A_\pm&=-\frac{f}{2}k^A_\pm.
\end{subalign}
In this way, the transition function from one basis into the other, namely \eref{Udef}, remains $\delta_f$-invariant,
\begin{equation}
\delta_f\ou{U}{A}{B}=0.
\end{equation}
It is now also useful to split $f:\mathcal{N}\rightarrow\C$ into its real and imaginary parts $\omega$ and $\phi$ respectively,
\begin{equation}
f=\omega+\I\phi.
\end{equation}
The variation of the composite fields (\ref{pmtetra}--\ref{pmSigma}) yields,
\begin{subalign}
\delta_f\vou{\pm}{e}{AA'}{a}&=\omega\,\vou{\pm}{e}{AA'}{a},\label{Cvar1}\\
\delta_f\vou{\pm}{\eta}{A}{ab}&=\frac{3\omega-\I\phi}{2}\,\vou{\pm}{\eta}{A}{ab},\label{Cvar2}\\
\delta_f\vou{\pm}{\Sigma}{AB}{ab}&=2\omega\,\vou{\pm}{\Sigma}{AB}{ab}.\label{Cvar3}
\end{subalign}
The variation of the difference tensor \eref{difftensor}, on the other hand, is given by
\begin{equation}
\delta_f\Delta^{AB}=k^{(A}\ell^{B)}\di f.
\end{equation}
It is now easy to evaluate the variation of the dressed Chern\,--\,Simons action \eref{CSdressed}, with respect to such transformations. A short calculation gives
\begin{align}
\nonumber\delta_fS_{\mathrm{CS}}&=\frac{1}{2}\int_{\mathcal{N}}\bigg[-F_{AB}k^A\ell^B\wedge \di f+\di f\, k_A\ell_B\wedge\ou{\Delta}{A}{C}\wedge\Delta^{CB}=\\
&=\frac{1}{2}\int_{\mathcal{N}}\Big[-F_{AB}k^A\ell^B\wedge \di f+\di f\wedge Dk_A\wedge D\ell^A\Big].
\end{align}
The integrand is a total derivative. This can be seen as follows. Consider first the one-form
\begin{equation}
a=k_AD\ell^A\in T^\ast\mathcal{N},
\end{equation}
which is  the abelian part of the Ashtekar connection \eref{difftensor}. Its exterior  derivative is given by
\begin{equation}
\di a=D k_A\wedge D \ell^A-F^{AB}k_A\ell_B,
\end{equation}
such that the conformal variation of the Chern\,--\,Simons functional vanishes up to a boundary term,
\begin{equation}
\delta_fS_{\mathrm{CS}}=\frac{1}{2}\int_{\mathcal{N}}\di f\wedge\di a=\frac{1}{2}\int_{\partial\mathcal{N}}f\di a.\label{CCS}
\end{equation}
Consider then the conformal variation of the entire gauge-fixed action \eref{gfixedactn} for boundary conditions $f|_{\partial\mathcal{N}}=0$. From (\ref{Cvar1}--\ref{Cvar3}) and \eref{CCS}, we immediately find,
\begin{align}\nonumber
\delta_fS_{\mathrm{gf}}=\frac{3\I}{8\pi\Lambda G}\int_{\mathcal{N}}\bigg[&\frac{2\Lambda}{3}\omega\big(\eta^+_A\wedge D\ell^A_+-\eta^-_A\wedge D\ell^A_-\big)+\frac{\Lambda}{6}\big(\eta^+_A\ell^A_+-\eta^-_A\ell^A_-\big)\wedge \di f+\\
&-2\big(B^+_{AB}\Sigma_+^{AB}-B^-_{AB}\Sigma_-^{AB}+\CC\big)\wedge\di\omega\bigg].
\end{align}
where $\approx$ denotes again equality up to terms that vanish \emph{on-shell}. The first term vanishes thanks to the gluing conditions \eref{match1} and \eref{connctnmatch1}, and the second term vanishes due to the matching conditions for the codyad $(m_a,\bar{m}_a)$, namely \eref{gluecond1}. We are therefore left with the last term that implies a condition on the spin $(1,0)$ part of the auxiliary $B_{\alpha\beta}^\pm$-fields, namely that the corresponding two-form is closed,
\begin{equation}
\di\big(B^+_{AB}\Sigma^{AB}_+-B^{-}_{AB}\Sigma_-^{AB}+\CC\big)=0.\label{Bexact}
\end{equation}

In summary, the field equations derived from the gauge-fixed boundary action \eref{gfixedactn} are (i) the torsionless conditions \eref{torsless} for the pull back of the spin connection to the boundary, derived from the variation of the auxiliary $B^\pm_{\alpha\beta}$-fields, (ii) the curvature constraints \eref{bndryF}, which are equivalent to the constraint equations (\ref{psi1eq}--\ref{psi3eq}) and (\ref{evolv1}--\ref{evolv5}) of general relativity on a null hypersurface and (iii) the supplementary conditions \eref{Bterms} and \eref{Bexact} for the auxiliary $B^\pm_{\alpha\beta}$-fields.

\section{Boundary phase space and quasi-local amplitudes}\label{sec4}
\subsection{Torsionless condition as a gauge fixing term}\label{sec4.1}
\noindent  In the last section, we saw that the field equations for the gauge-fixed boundary action \eref{gfixedactn} impose the constraint equations of general relativity on a null hypersurface: the variation of the auxiliary $B^\pm_{\alpha\beta}$-fields impose the torsionless condition on the boundary, while the boundary equations of motion \eref{bndryF} imply that the curvature of the self-dual connection is compatible with the pull-back to the null boundary of the Riemann curvature tensor for a solution of the $\Lambda$-vacuum Einstein equations. So far, we have only defined the action \eref{gfixedactn} at a gauge-fixed level. The question is then how to unfreeze the gauge fixing conditions and distinguish spurious gauge directions from actual physical degrees of freedom. This question is less trivial than it seems. Clearly, the orbits of the $U(1)$ gauge transformations \eref{U1gauge} and dilations \eref{dilat} should be treated as unphysical gauge directions on phase space, and we may therefore simply remove the auxiliary $B$-field from the action. But there may exist further such gauge directions hiding in the boundary action \eref{gfixedactn}. The main justification and expectation for this viewpoint is that the auxiliary $B^\pm_{\alpha\beta}$-fields that impose the torsionless condition fall completely out of the set of evolution and constraint equations on the null hypersurface, see \eref{bndryF} and \eref{10torsion} and \eref{11torsion}. This is very reminiscent of gauge fixing terms: auxiliary fields that impose gauge fixing conditions decouple from the equations of motion for gauge-invariant variables. Its is therefore tempting to think that the fundamental gauge-unfixed action is simply given by the following expression,
\begin{align}
\nonumber S\big[(k^A_\pm,\ell^A_\pm),\vou{\pm}{A}{A}{Ba},\psi_{ABC},&m_a|h\big]=\\
\nonumber=\frac{3\I}{8\pi\Lambda G}\int_{\mathcal{N}}\bigg[&+\frac{\Lambda}{3}\eta^+_A\wedge D\ell^A_+ -\frac{1}{2}\mathrm{Tr}\big[F^+\wedge\Delta^+-\frac{1}{3}\Delta^+\wedge\Delta^+\wedge\Delta^+\big]+
 \\
&\nonumber- \frac{\Lambda}{3}\eta^-_A\wedge D\ell^A_-+\frac{1}{2}\mathrm{Tr}\big[F^-\wedge\Delta^--\frac{1}{3}\Delta^-\wedge\Delta^-\wedge\Delta^-\big]+\\
&+\psi_{ABC}\,\eta^A_+\wedge\big[UDU^{-1}\big]^{BC}\bigg]\equiv\int_{\mathcal{N}}\big(\mathcal{L}_+-\mathcal{L}_-\big),\label{Sactndef}
\end{align}
where, as in \eref{kdef} and \eref{etadef}, the spinor-valued two-form $\eta^\pm_{Aab}$ is the composite field,
\begin{equation}
\eta^\pm_A=-\big(\ell_A^\pm\di u+k_A^\pm m)\wedge \bar{m},
\end{equation}
and $\ou{U}{A}{B}$ denotes the transition function \eref{Udef} between the two spin frames. The equivalence class $h=[m_a]$ is again to be treated as an external source at the boundary. If this viewpoint is adopted, the torsionless condition is a mere gauge-fixing term for the complex-valued action \eref{Sactndef}. Initial data for general relativity on a null hypersurface would be  recovered  by the restoration to only those gauge sections, where the torsionless condition holds. The entire procedure is clearly reminiscent of the original self-dual formulation of loop quantum gravity, where the reality conditions can be recast into gauge fixing conditions for the complex theory, see e.g.\ \cite{BodendorferNewVarI,komplex1}.

\subsection{Doubled phase space and the structure of the transition amplitudes}\label{sec4.2}
\noindent Finally, let us briefly discuss what we could learn from this boundary field theory for quantum gravity in the bulk. Consider first the structure of the classical phase space. The pre-symplectic potential of the boundary field theory is obtained from the first variation of the boundary action \eref{Sactndef}. This one-form on field space is a sum of terms intrinsic to either side of the null hypersurface,
\begin{equation}
\Theta_{S^2}=\Theta^+_{S^2}-\Theta^-_{S^2}.
\end{equation}
The symplectic potential for each individual part of the component system is now given by the two-dimensional integral
\begin{equation}
\Theta^{\pm}_{S^2}=-\frac{3\I}{16\pi\Lambda G}\oint_{S^2_{u_o}}\bigg[\Delta^\pm_{AB}\bbvar{d}A^{AB}_\pm+F^\pm_{AB}\big(k^A_\pm\bbvar{d}\ell^{B}_\pm-\ell^A_\pm\bbvar{d}k^B_\pm\big)\bigg],\label{ThetaS}
\end{equation}
on an arbitrary $u=u_o$ cross-section of the null hypersurface. Notice that this is a \emph{two-dimensional} integral. The reason is that we are dealing with a \emph{three-dimensional} boundary field theory, whose field equations are the constraint equations\footnote{The constraints are given by the pull back of the torsionless condition $\nabla\vo{4}{e}{AA'}=0$  and the curvature  constraint $\vou{4}{F}{A}{B}=\frac{\Lambda}{3}\vou{4}{\Sigma}{A}{B}+\ou{\Psi}{A}{BCD}\vo{4}{\Sigma}{CD}$ to the null boundary.} of general relativity on a null hypersurface. 

Performing a Legendre tarnsformation, we obtain the $u$-dependent Hamiltonian, which splits into a sum of terms from either side of the hypersurface,
\begin{subalign}
H[h]&=H^+[h]-H^-[h],\label{Hamdef1}\\
H^\pm[h]&=\int_{S^2_u}\Theta^\pm_{S^2}(\mathcal{L}_{\partial_u})-\partial_u\hook\mathcal{L}^\pm,\label{Hamdef2}
\end{subalign}
where the notation $H^\pm[h]$ indicates that the conformal class $h=[m_a]$ appears as an external source at the boundary, see \eref{boundrycond}. In general, the Hamiltonian will be therefore time dependent. For the present argument, the concrete expressions of the boundary Hamiltonian $H^\pm$ in terms of phase space variables is of limited importance. What matters is that the total Hamiltonian $H=H^+-H^-$ vanishes as a constraint, if the gluing conditions (\ref{match1}--\ref{match3}, \ref{gaugecond1}, \ref{gaugecond2}) are  satisfied. The action \eref{Sactndef} describes, therefore, a prototypical example of a timeless or relativistic system \cite{PhysRevD.42.2638, rovelli}, where an external clock can only arise once we split the system into its component parts. Indeed, physical states of the coupled system will be of the following general form 
\begin{equation}
\Psi(q^+,q^-)=\sum_{nmm'}c_{nmm'}\psi_{nm}(q^+)\psi_{nm'}(q^-),\quad \sum_{nmm'}|c_{nm'}|^2=1,\label{physstates}
\end{equation}
where the edge states $\psi_{nm}\in\mathcal{H}_{S^2}$ are eigen vectors of the quasi-local Hamiltonian,\footnote{A description of such edge states from the perspective of loop quantum gravity in the bulk has been given in \cite{Wieland:2017cmf}.  The quantum number $m$ labels the potential degeneracy of $H^\pm[h]$.}
\begin{equation}
H^\pm\psi_{nm} = E_{n} \psi_{nm}.
\end{equation}
There is, therefore, no preferred clock available at the boundary. One possibility to reintroduce an external time variable and study ordinary transition amplitudes is to simply split the system into its two component parts \cite{PhysRevD.42.2638}. For each individual subsystem, we will now have an ordinary Schrödinger evolution along the generators of the null hypersurface, and the resulting boundary transition amplitudes describe the evolution between two consecutive cross-sections of the null hypersurface,
\begin{equation}
A_{\mathcal{N}}\big(\bar{\psi}_{1},\psi_{0},h\big)=\big\langle \psi_1\big|\mathrm{Texp}\Big(-\I\displaystyle{\int_{u_0}^{u_1}}\di u\,H^+[h]\Big)\big|\psi_0\big\rangle.\label{bampl}
\end{equation}
Notice then that these amplitudes not only depend on the \emph{in} and \emph{out} edge states $\psi\in\mathcal{H}_{S^2}$, but that there is also a functional dependence on additional data, namely the equivalence class $h=[m_a]$, which characterises the gravitational flux across the null hypersurface.\footnote{Up to an $u$-independent zero mode, the evolution equation \eref{evolv3} allows us to identify the transversal shear $\sigma_{(k)}$ with the equivalence class $h=[m_a]$. This choice would be closer to the treatment at null infinity, where the gravitational flux is encoded into the asymptotic shear $\sigma^0=\lim_{r\rightarrow\infty}r^2\sigma_{(k)}$ that characterises the two radiative modes at the full non-linear level \cite{AshtekarNullInfinity}.}  The boundary amplitudes \eref{bampl} may be viewed, therefore, as state vectors on an extended state space $\mathcal{H}=\mathcal{H}^\ast_{S^2}\otimes\mathcal{H}_{\rm{flux}}\otimes\mathcal{H}_{S^2}$, which is now assigned to the entire null hypersurface $\mathcal{N}$ and not just to a two-dimensional cross-section thereof. An analogue of the gravitational $S$-matrix may be then reconstructed by simply gluing two such amplitudes for expanding (contracting) null boundaries $\mathcal{N}_\pm$ along a common cross-section, thus obtaining, by the standard rules of quantum mechanics
\begin{equation}
S\big(\bar{\psi}_+,\psi_-,[\bar{h}_+],[h_-]\big)=\sum_{nm} \overline{A_{\mathcal{N}_+}(\bar{\psi}_{nm},\psi_+,[h_+])}A_{\mathcal{N}_-}(\bar{\psi}_{nm},\psi_-,[h_-]).
\end{equation}
Notice that these amplitudes depend not only on the gravitational data $[h_\pm]$  at the null hypersurfaces $\mathcal{N}_\pm$, but they also depend on additional corner data $\psi_\pm\in\mathcal{H}_{S^2}$ at the endpoints of the boundary $\mathcal{N}=\mathcal{N}_+\cup\mathcal{N}_-$. The main conceptual difference to the usual perturbative $S$-matrix approach concerns the choice of boundary states against which these amplitudes are evaluated.  In our framework, the boundary states will live on a finite null boundary rather than at future (past) null infinity. The limit to asymptotic boundaries can then only be obtained \emph{within} the quantum theory, by selecting a limit of edge states $\psi\in S^2$ describing ever larger configurations of the two-dimensional $u=\mathrm{const}.$ slices.  
\section{Conclusion}
\noindent Let us summarise. The first part of the developed a classical boundary field theory, whose equations of motion are the constraint equations of general relativity on a null hypersurface. The basic strategy was to couple the analogue of the Gibbons\,--\,Hawking\,--\,York boundary term on a null hypersurface, namely \eref{bndryterm}, to the dressed Chern\,--\,Simons term \eref{CSdressed}. The resulting boundary action is a generating functional for the pull-back of the self-dual part of the De Sitter curvature $\mathcal{F}_{\alpha\beta}=F_{\alpha\beta}-\frac{\Lambda}{3}e_\alpha\wedge e_\beta$ on a null hypersurface, see \eref{test1}. This coupled boundary action by itself cannot describe gravitational radiation, because the pull-back of the Weyl tensor would vanish at the boundary. To capture the two propagating degrees of freedom, we then found it necessary to double the boundary field content $(p,q)\rightarrow(p^\pm,q^\pm)$ and impose additional gluing conditions (\ref{matchcond1}--\ref{matchcond4}) between the two sectors of the theory. At the level of the boundary action, the gluing conditions are imposed via a spin $(\tfrac{3}{2},0)$ Lagrange multiplier $\psi_{ABC}$ that represents the only visible components of the Weyl tensor at the null hypersurface, namely $\psi_{ABC}=\Psi_{ABCD}\ell^D$. Indeed, from the perspective of the boundary field theory, the curvature spinor $\Psi_0=\Psi_{ABCD}k^Ak^Bk^Ck^D$ is inaccessible, because it cannot be obtained from the pull-back  $\vt{4}{R}^{a}{}_{b\underleftarrow{cd}}$ of the Riemann curvature two-form to the null hypersurface.\footnote{If we extend the transversal null direction $k^a$ into a geodesic that enters the bulk, we can introduce an affine Bondi coordinate $r$ such that $k^a=\partial_r^a$ in which case $\Psi_0$ is determined by the radial evolution equation $\Psi_0=\frac{\di}{\di r}\bar{\sigma}_{(k)}+\vartheta_{(k)}\bar{\sigma}_{(k)}$.}

The last part of the paper was more conceptual in nature. In section \hyperref[sec4.1]{section 4.1}, we reconsidered Ashtekar's original self-dual formulation of loop gravity \cite{newvariables, ashtekar} and argued that the boundary action \eref{gfixedactn} can be considerably simplified by treating the torsionless equation at the boundary as a gauge fixing condition for a more general boundary field theory for a complex connection. This was justified by the observation that the auxiliary fields $B^\pm_{\alpha\beta}$ that impose the torsionless conditions fall out of the equations of motion for the tetrad and connection at the null boundary, see e.g.\ \eref{Bterms} and (\ref{Aplusvar}--\ref{bndryF}). This is reminiscent to what happens for gauge fixing conditions in field theory: corresponding Lagrange multipliers do not enter the field equations for gauge invariant variables.\footnote{The value of the Lagrange multiplier $\xi$ that imposes the e.g.\ Lorentz gauge condition $\nabla_aA^a=0$ is completely undetermined and does not enter the field equations $\nabla_a F^{ab}=0$ for the gauge invariant observables.} The entire approach resembles, therefore, the original self-dual formulation of loop gravity \cite{ashtekar}, where the reduction to a real slice in phase appears as a mere gauge choice after having solved the equations of motion for the now complex theory, see \cite{BodendorferNewVarI,komplex1} for a similar gauge unfixing procedure in the bulk \cite{Vytheeswaran:1994np}. Finally, we explained the relation to holography and gave a sketch for how to generate transition amplitudes for quantum general relativity in the bulk from transition amplitudes at the boundary. The basic idea was that  transition amplitudes for edges states $\psi_{\text{in}}, \psi_{\text{out}}\in\mathcal{H}_{S^2}$ at two consecutive cross sections of the null boundary depend parametrically on the intermediate flux,\footnote{By integrating \eref{evolv3}, we can identify the gauge equivalence class $[h]$ with the transversal shear $\sigma_{(k)}$, which encodes the gravitational radiation at $\mathcal{I}^\pm$ in an $r\rightarrow\infty$ limit.} which is encoded into the gauge equivalence class $[m_a]$. The transition amplitudes of the boundary field theory may therefore be interpreted as vectors on an enlarged state space $\mathcal{H}=\mathcal{H}^\ast_{S^2}\otimes\mathcal{H}_{\rm{flux}}\otimes\mathcal{H}_{S^2}$, whose elements describe not only the gravitational edge states at $\partial\mathcal{N}$ but also the gravitational flux across the null boundary $\mathcal{N}$ in between. The  construction of such boundary states $\Psi\in\mathcal{H}=\mathcal{H}^\ast_{S^2}\otimes\mathcal{H}_{\rm{flux}}\otimes\mathcal{H}_{S^2}$ from the transition amplitudes of an auxiliary boundary field theory would be a concrete realisation of the quasi-local holographic principle, where the quantum states of the theory are defined at finite boundaries, and the limit to asymptotic boundaries is obtained only within the quantum gravity, by defining e.g.\ an appropriate family of boundary coherent states. A concrete realisation of this construction for euclidean gravity in three dimensions can be found in e.g.\ \cite{Bonzom:2015ans,Dittrich:2018xuk,Dittrich:2017hnl,Wieland:2018ymr}.

\paragraph{Acknowledgments} I am indebted to the researchers at Perimeter Institute for most fruitful discussions and feedback and I would like to thank Laurent Freidel in particular. This research emerged from a visit to Penn State and I thank Abhay Ashtekar and Eugenio Bianchi for most helpful comments, feedback and discussions during the initial phase. In addition, I thank Simone Speziale for comments on a first draft of the paper. Research at Perimeter Institute is supported by the Government of Canada through the Department of Innovation, Science and Economic Development and by the Province of Ontario through the Ministry of Research and Innovation.

\providecommand{\href}[2]{#2}\begingroup\raggedright\endgroup

\end{document}